\documentclass{pasj01}
\Received{$\langle$reception date$\rangle$}
\Accepted{$\langle$acception date$\rangle$}
\Published{$\langle$publication date$\rangle$}

\begin{document}

\title{Observational Study on the Fine Structure and Dynamics of a Solar Jet.\\
I. Energy Build-Up Process around a Satellite Spot.}
\author{Takahito \textsc{Sakaue}\altaffilmark{1}, Akiko \textsc{Tei}\altaffilmark{1}, Ayumi \textsc{Asai}\altaffilmark{1}, Satoru \textsc{Ueno}\altaffilmark{1}, Kiyoshi \textsc{Ichimoto}\altaffilmark{1}, \& Kazunari \textsc{Shibata}\altaffilmark{1}}%
\altaffiltext{1}{Kwasan and Hida Observatories, Kyoto University, Kyoto, Japan}
\email{sakaue@kwasan.kyoto-u.ac.jp}

\KeyWords{Sun: flares  --- Sun: granulation --- Sun: magnetic fields --- sunspots}

\maketitle

\begin{abstract}
We report a solar jet phenomenon associated with successive flares on November 10th 2014. These explosive events were involved with the satellite spots' emergence around a $\delta$-type sunspot in the decaying active region NOAA 12205. The data of this jet was provided by {\it Solar Dynamics Observatory} ({\it SDO}), X-Ray Telescope (XRT) aboard {\it Hinode}, {\it Interface Region Imaging Spectrograph} ({\it IRIS}) and Domeless Solar Telescope (DST) at Hida Observatory, Kyoto University. These plentiful data enabled us to present this series of papers to discuss the entire processes of the observed phenomena including the energy storage, event trigger, and energy release. In this paper, we focus on the energy build-up and trigger phases, by analyzing the photospheric horizontal flow field around the active region with an optical flow method. The analysis reveals the following three. (i) The observed explosive phenomena involved three satellite spots, the magnetic fluxes of which successively reconnected with their pre-existing ambient fields. (ii) All of these satellite spots emerged in the moat region of a pivotal $\delta$-type sunspot, especially near its convergent boundary with the neighboring supergranules or moat regions of adjacent sunspots. (iii) Around the jet ejection site, the positive polarities of satellite spot and adjacent emerging flux encountered the global magnetic field with negative polarity in the moat region of the pivotal $\delta$-type sunspot, and thus the polarity inversion line was formed along the convergent boundary of the photospheric horizontal flow channels.
\end{abstract}

\section{Introduction}
\label{sec:Introduction}
The evolution of solar active region leads to the explosive phenomena in the solar atmosphere such as flares and jets. Those phenomena are identified as the transient energy release process involving a large amount of magnetic energy which is stored in the coronal magnetic field. Because it is difficult to measure the coronal magnetic field directly, the energy build-up process in the corona has been deduced from the distribution of the concentrated magnetic fluxes on the photosphere and from their relative motions such as the shear or convergent motions
(\cite{2000ApJ...540..583S},
\cite{2000PASJ...52..337I},
\cite{2001ApJ...552..833M},
\cite{2008PASJ...60.1181M},
\cite{2010PASJ...62..921K}).
There are many numerical experiments for the explosive phenomena in order to relate them with the photospheric magnetic field 
(\cite{2009ApJ...691...61P},
\cite{2012ApJ...760...31K}).
\par
Recently, owning to the development of the data-driven simulations and the coronal field extrapolation methods, it became possible to investigate the energy build-up process in the coronal magnetic field above the active region, no matter how complicatedly the magnetic fluxes distribute on the photosphere.
(\cite{2013A&A...559A...1S},
\cite{2013ApJ...773..128T},
\cite{2015ApJ...801...83C},
\cite{2016NatCo...711522J}).\par
Such an evolution of magnetic configuration in the solar atmosphere is eventually attributed to the flux emergence process in which the magnetic energy is transported from the convection zone to the solar atmosphere. From this perspective, 
\cite{1977ApJ...216..123H} 
elaborated the self-consistent scenario in which the various explosive phenomena are caused by the interaction between the pre-existing coronal field and an emerging flux. This scenario has been realized in many numerical experiments (\cite{1995Natur.375...42Y}, 
\cite{2013ApJ...771...20M}). 
Moreover, it is notable that their simulations have established the unified model on the dynamics of the explosive phenomena such as jet and flare, based on the ubiquity of the magnetic reconnection process (\cite{1988ApJ...330..474P}, \cite{2007Sci...318.1591S}). According to this unified model (\cite{1999Ap&SS.264..129S}), the characteristic temperature and velocity of an explosive phenomenon are determined by where the magnetic energy is released through the magnetic reconnection, which is possible to be triggered anywhere as long as in the magnetized plasma. In fact, the reconnection in the chromosphere produces the cool jet such as H$\alpha$ jet, as reproduced by \cite{2013PASJ...65...62T}, and that in the corona follows the hot jet such as EUV jet or X-ray jet, as reproduced by \cite{2008ApJ...683L..83N} and \cite{2004ApJ...614.1042M}. Note that the physical mechanisms driving those diverse jets are different from each other; for example, the cool jet reproduced by \cite{1995Natur.375...42Y} is driven by the Lorentz force, while that of \cite{2013PASJ...65...62T} are driven by the interaction between the slow-mode shock and the transition layer. The hot jets reproduced by \cite{2008ApJ...683L..83N} and \cite{2004ApJ...614.1042M} are driven by the Lorentz force and pressure gradient force, respectively. It is notable that those different driving mechanisms are self-consistently determined to each jet according to the site of the reconnection between the emerging flux and the ambient field.\par

As mentioned above, it is well understood how an individual flux emergence associates the various explosive phenomena by its interaction with the pre-existing coronal field. On the other hand, an actual active region, especially flare-productive one, comprises various concentrated magnetic constituents with wide range of spatial scale, such as sunspots, pores, and faculare arranged in plages and enhanced network around the time of its maximum development \citep{2000ssma.book.....S}. The active region experiences the frequent emergences of magnetic fluxes and the continuous disintegrations of them. As a result of such a self-organization of an entire active region, the explosive phenomena in it are often characterized by the evolutionary phase of the active region. For example, the jet activity has been related to the earliest active region \citep{1993ASPC...46..507K}, light bridge \citep{2001ApJ...555L..65A}, moving magnetic features from a sunspot (\cite{2007ApJ...656.1197B}, \cite{2012ApJ...752...70U}), or satellite spots (\cite{1968IAUS...35...77R}, \cite{1972SoPh...25..141R}, \cite{1996ApJ...464.1016C}, \cite{1997SoPh..173..319H}, \cite{1998SoPh..178..379S}, \cite{1998ApJ...499..898I}, \cite{2002ApJ...574.1074S}, \cite{2014ChA&A..38...65W}, \cite{2015ApJ...815...71C}). These observational reports suggest the universal relationship between the flare or jet activity and the evolution of an entire active region. That is also important for the space weather or for the studies on the flare-activity of the stellar active region.\par

Nevertheless, because it is still challenging to theoretically address the whole evolutionary process of an actual active region from emergence to decay, the current knowledge on such a relationship is lack of self-consistency and far from versatile. The satellite spots, for example, have been paid attention to as the activator of the evolved active region, but there remain questions on their formation and connectivity with the pivotal sunspot accompanied by them. Further studies are needed to comprehend the process from the birth of the satellite spots to the explosive phenomena associated with them, in terms of an evolutionary history of the entire active region.\par

In this paper, therefore, we present a case study on a jet phenomenon, which was related to the evolution of a satellite spot in the active region NOAA 12205. It is notable that the jet was associated with successive flares (C3.6 and C5.4 class), especially the latter one. These overall explosive phenomena started at 23:26UT on November 10th 2014 and were extinguished approximately in an hour.\par
Owning to the continuous full-disk observation by {\it Solar Dynamics Observatory} ({\it SDO}; \cite{2012SoPh..275....3P}) including Atmospheric Imaging Assembly (AIA; \cite{2012SoPh..275...17L}) and Helioseismic and Magnetic Imager (HMI; \cite{2012SoPh..275..207S}), we are enabled to investigate those explosive phenomena in terms of their energy build-up, event trigger, and energy release phases. \par
Fortunately, this jet were also observed by the spectrograph of Domeless Solar Telescope in Hida Observatory, Kyoto University \citep{1985MmKyo..36..385N}. The spectral analysis results will be presented in Sakaue et al. 2017 (in preparation; hereafter Paper II) and the detailed dynamics of the observed jet in the energy release process will be discussed there. Thus, this paper focus on the energy build-up and event trigger phases. \par

We summarize the observed features of the jet and flares in section \ref{sec:Observation}, and explain the analysis process in section \ref{sec:Analysis}. The above-mentioned analysis results are described in section \ref{sec:Analysis} in more detail, and are referred to for the discussion in section \ref{sec:Discussion}.

\begin{figure*}[tbp]
\begin{center}
\FigureFile(150mm,230mm){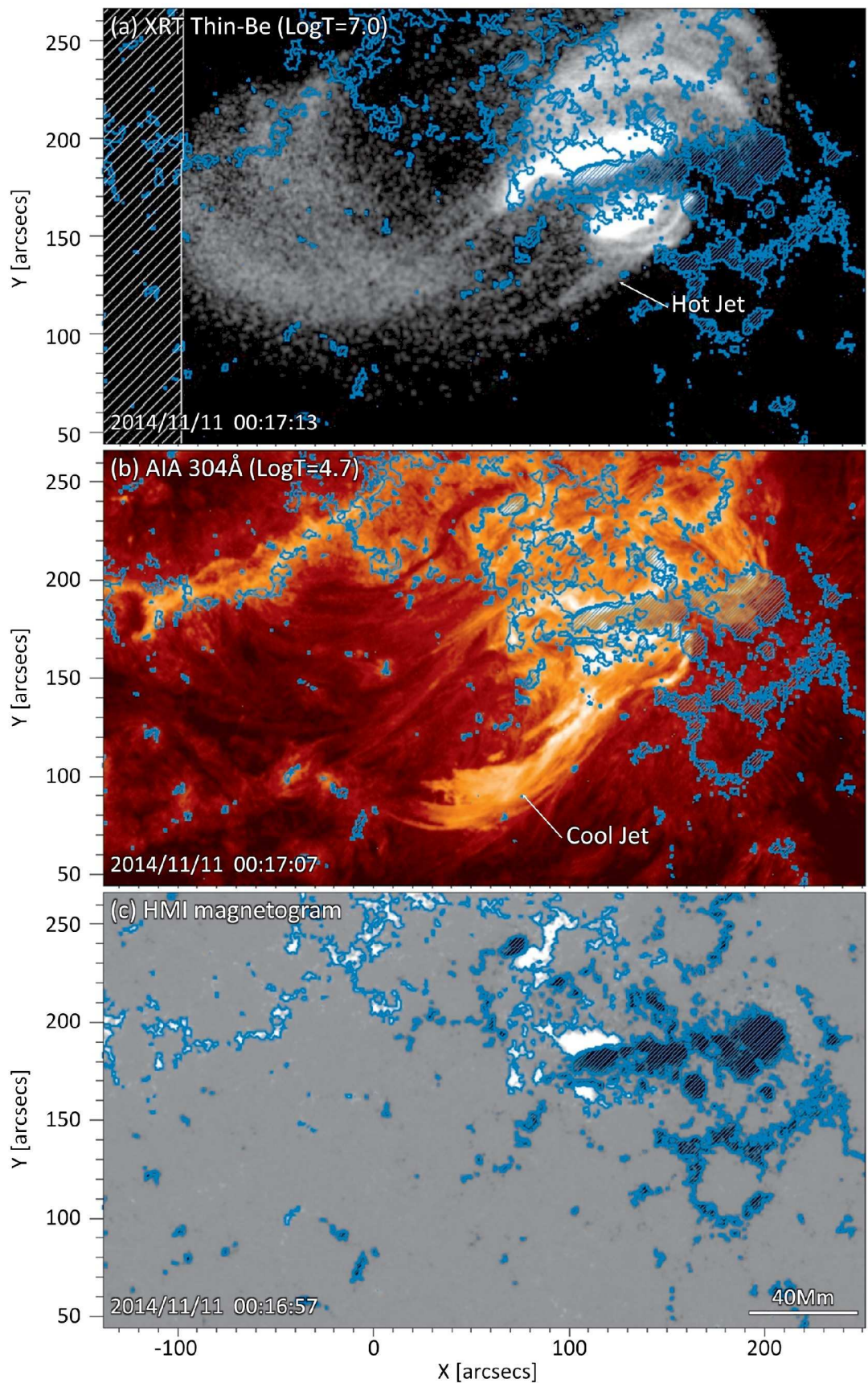}
\caption{The simultaneous snapshots of the observed jet in (a) {\it Hinode}/XRT images, (b)  {\it SDO}/AIA 304\AA, and (c) {\it SDO}/HMI line-of-sight magnetogram. The observed jet consisted of the cool ($\sim10^{4-5}$K) and hot ($\sim10^{6-7}$K) components. The contours in these images represent the regions where the absolute magnetic field strength is larger than 100G, and especially, the negative polarities are shaded.}
\label{fig:jet_multi}
\end{center}
\end{figure*}
\begin{figure*}[tbp]
\begin{center}
\FigureFile(90mm,45mm){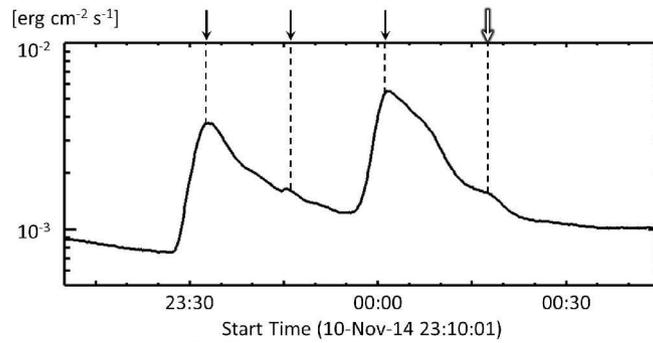}
\caption{The time variation of soft X-ray flux observed by GOES. The two peaks during the observed phenomena correspond to the successive flares (C3.6 and C5.4 class) in the same active region; NOAA 12205. The white arrow represents the time of the snapshots in Fig.\,\ref{fig:jet_multi} and each black arrow indicates that of panel (a), (b), or (c) of Fig.\,\ref{fig:events_aia094}, Fig.\,\ref{fig:events_aia1600} and Fig.\,\ref{fig:events_hmimag}}
\label{fig:goes_light_curve}
\end{center}
\end{figure*}
\begin{figure*}[tbp]
\begin{center}
\FigureFile(150mm,230mm){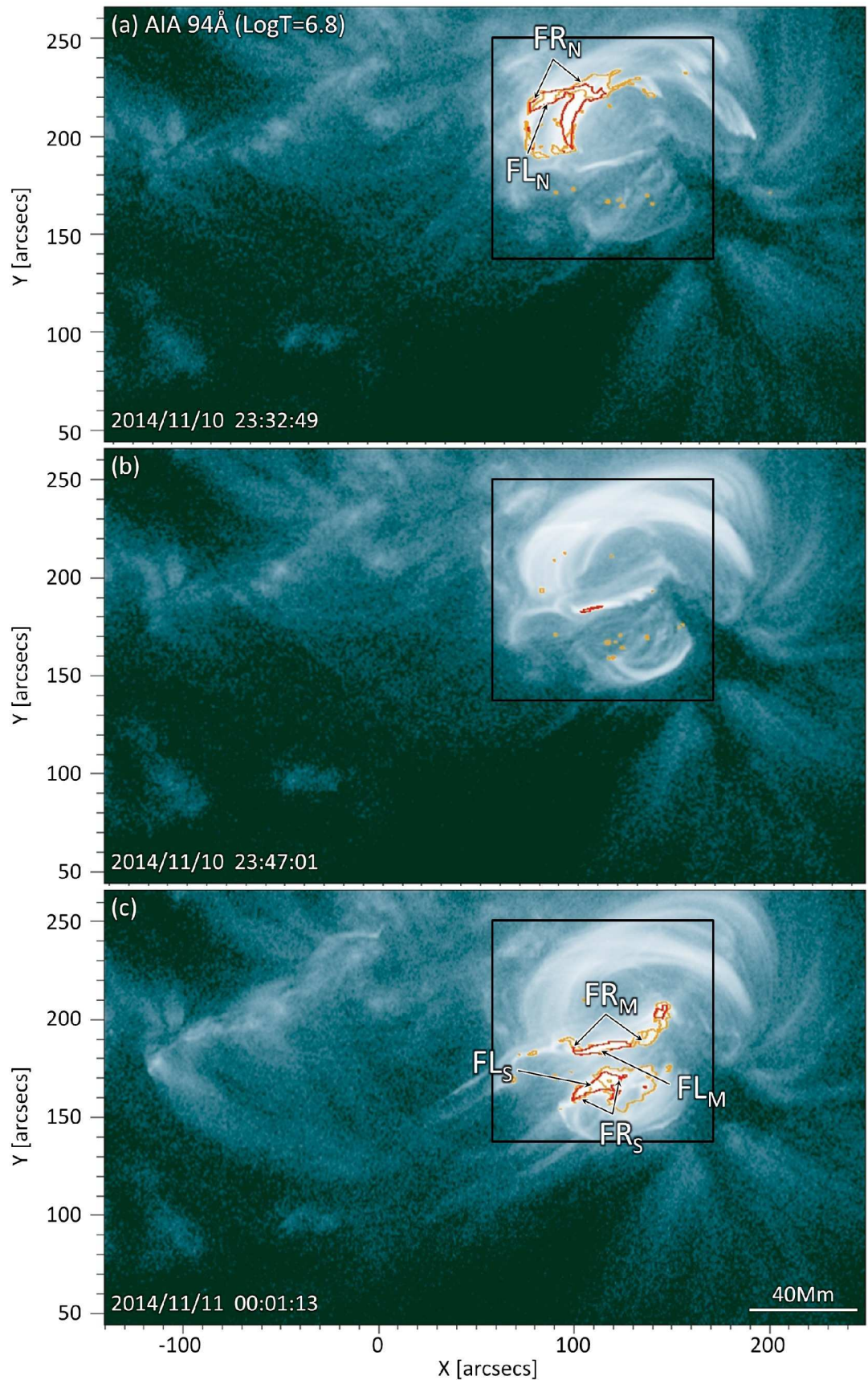}
\caption{The successive flares in AIA 94\AA\ images. The GOES soft X-ray flux at the time of each image is indicated with dash-dot lines in Fig.\,\ref{fig:goes_light_curve}. The red and yellow contours in those images represent the brightening features in AIA 94\AA\ and AIA 1600\AA\ images. Some of those red or yellow contours respectively correspond to the flare loops (labeled FL$_{\rm N}$, FL$_{\rm S}$, FL$_{\rm M}$) or flare ribbon pairs (labeled FR$_{\rm N}$, FR$_{\rm S}$, FR$_{\rm M}$).}
\label{fig:events_aia094}
\end{center}
\end{figure*}
\begin{figure*}[tbp]
\begin{center}
\FigureFile(150mm,230mm){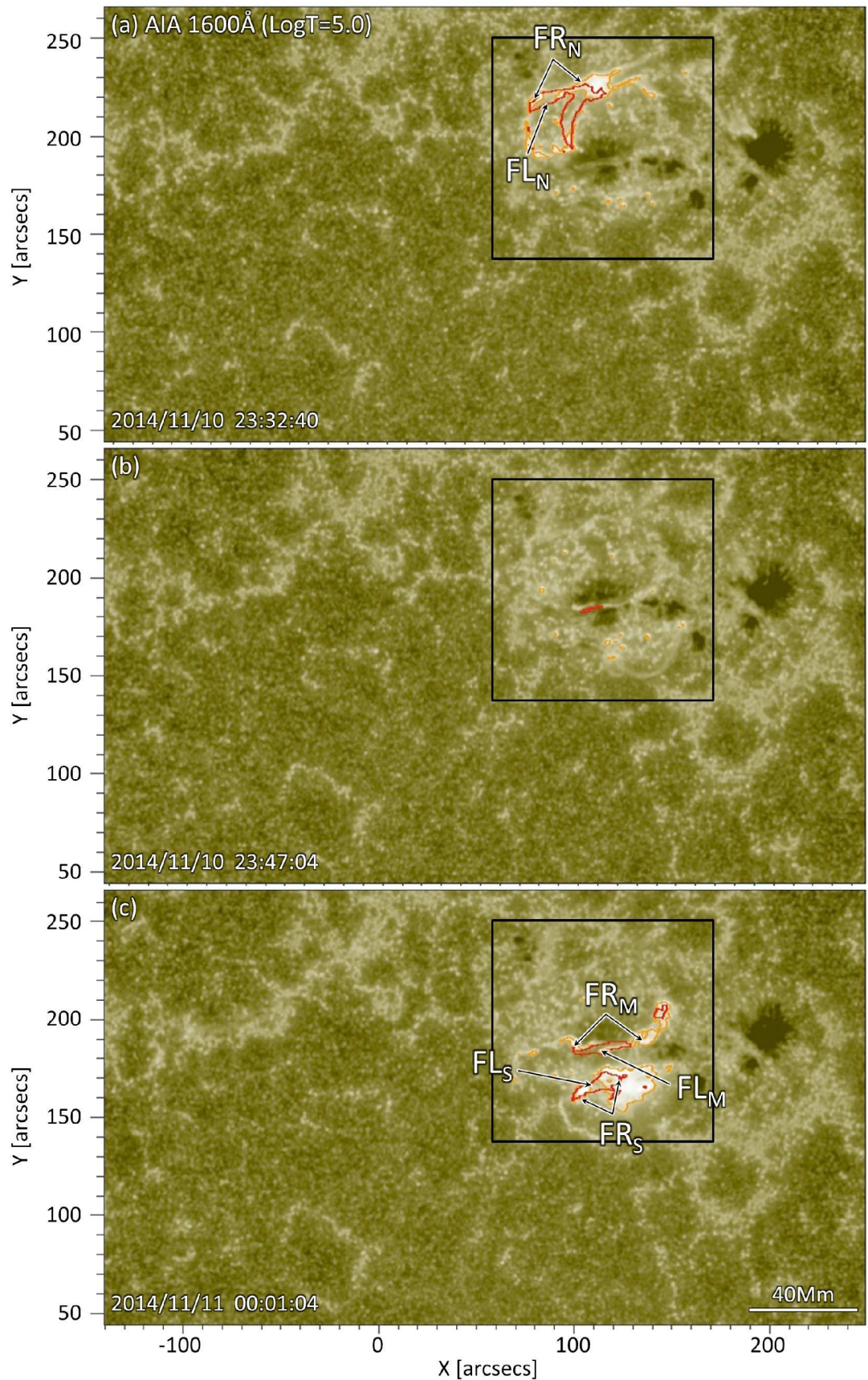}
\caption{The successive flares in AIA 1600\AA\ images. The field of view and times of these images are the same as those of Fig.\,\ref{fig:events_aia094}. Three pairs of flare ribbon FR$_{\rm N}$, FR$_{\rm M}$, and FR$_{\rm S}$ are shown in this figure.}
\label{fig:events_aia1600}
\end{center}
\end{figure*}
\begin{figure*}[tbp]
\begin{center}
\FigureFile(150mm,230mm){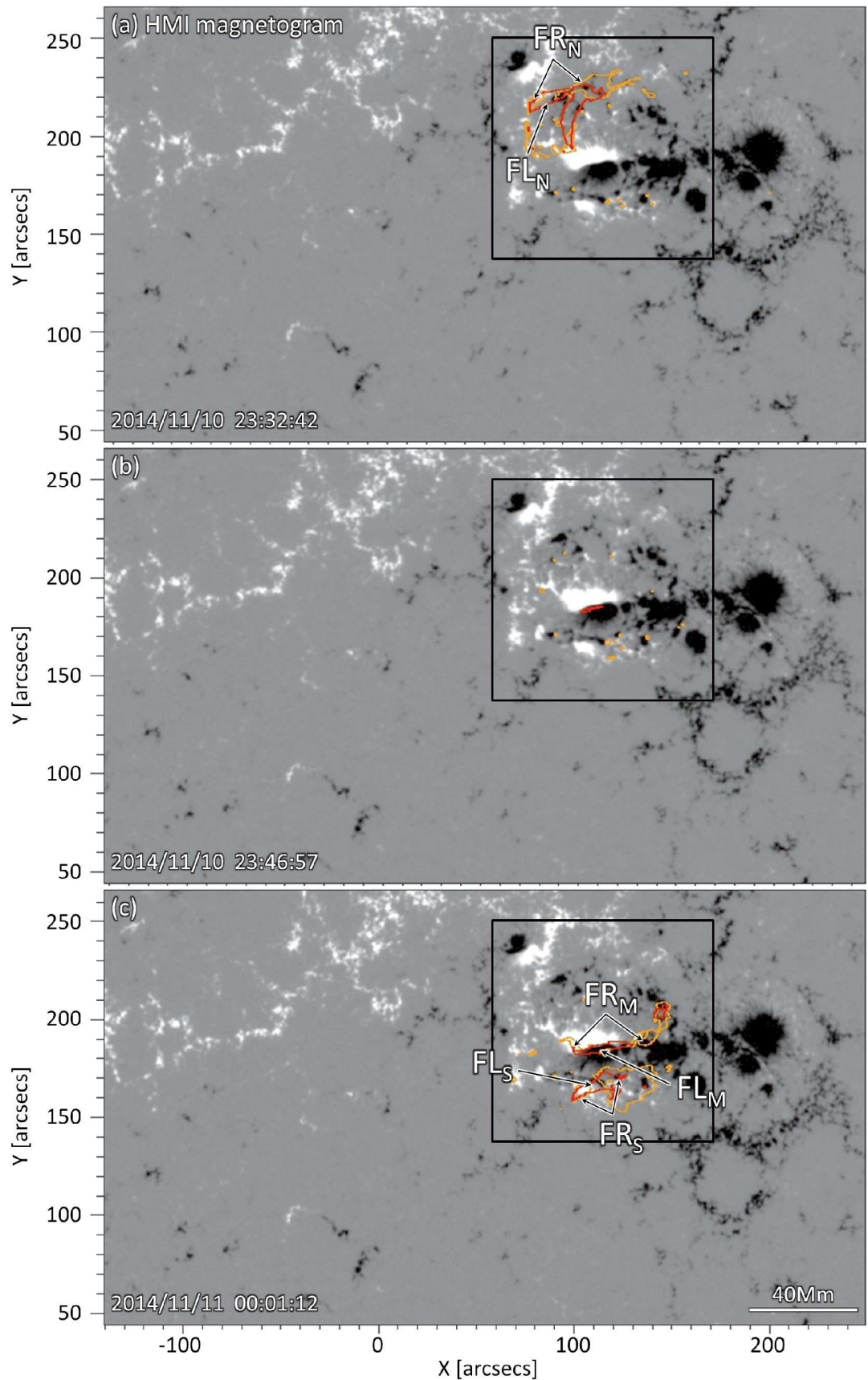}
\caption{The successive flares in HMI line-of-sight magnetograms. The field of view and times of these images are the same as those of Fig.\,\ref{fig:events_aia094}. The polarities of three pairs of flare ribbon FR$_{\rm N}$, FR$_{\rm M}$, and FR$_{\rm S}$ are shown in this figure.}
\label{fig:events_hmimag}
\end{center}
\end{figure*}
\begin{figure*}[tbp]
\begin{center}
\FigureFile(170mm,191mm){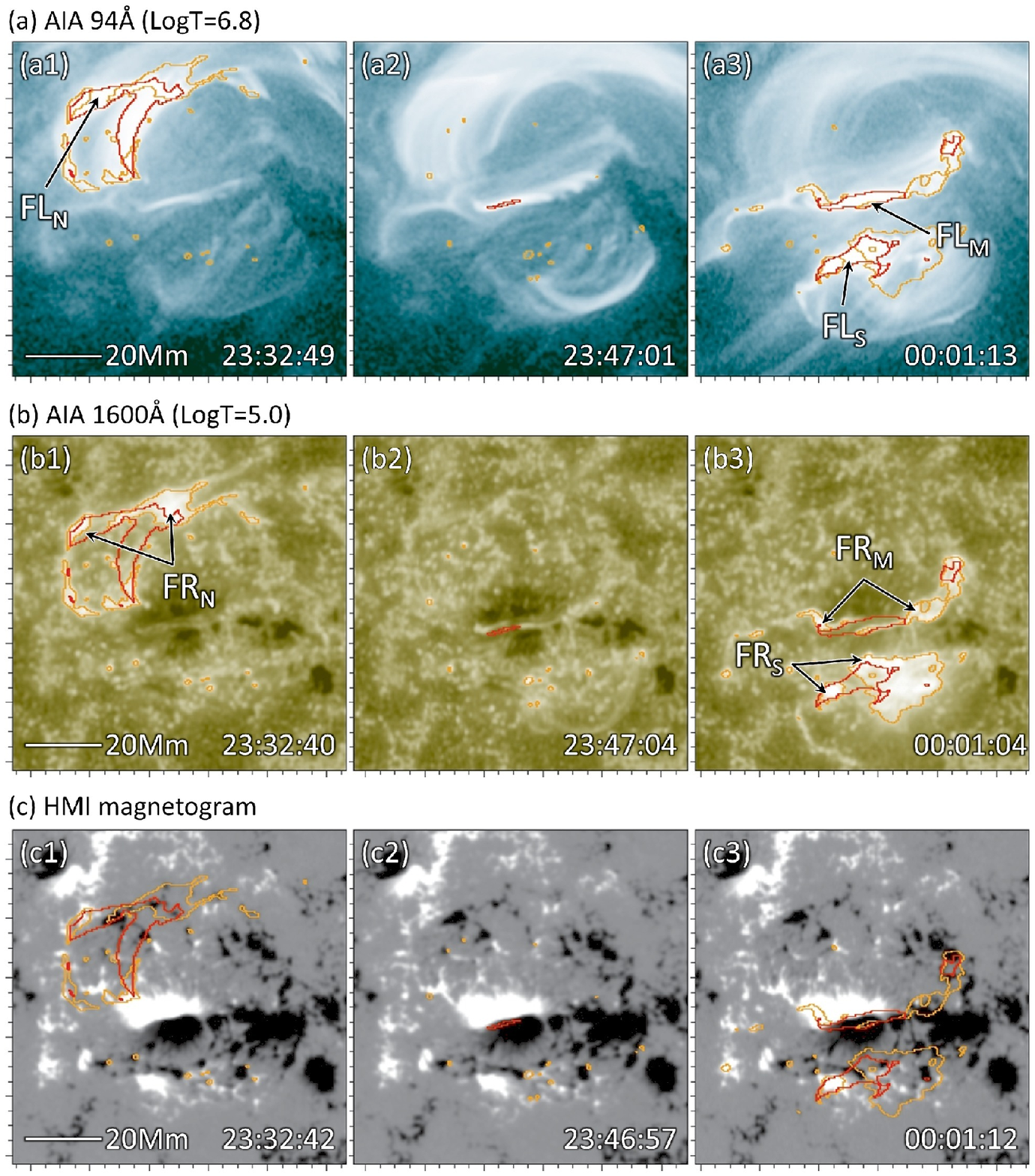}
\caption{The close-up views of black frames in Fig.\,\ref{fig:events_aia094}--\ref{fig:events_hmimag}. The successive flares in  (a) AIA 94\AA\ images, (b) AIA 1600\AA\ images, and in (c) HMI line-of-sight magnetograms.}
\label{fig:events_rb_fl}
\end{center}
\end{figure*}
\begin{figure*}[tbp]
\begin{center}
\FigureFile(170mm,112mm){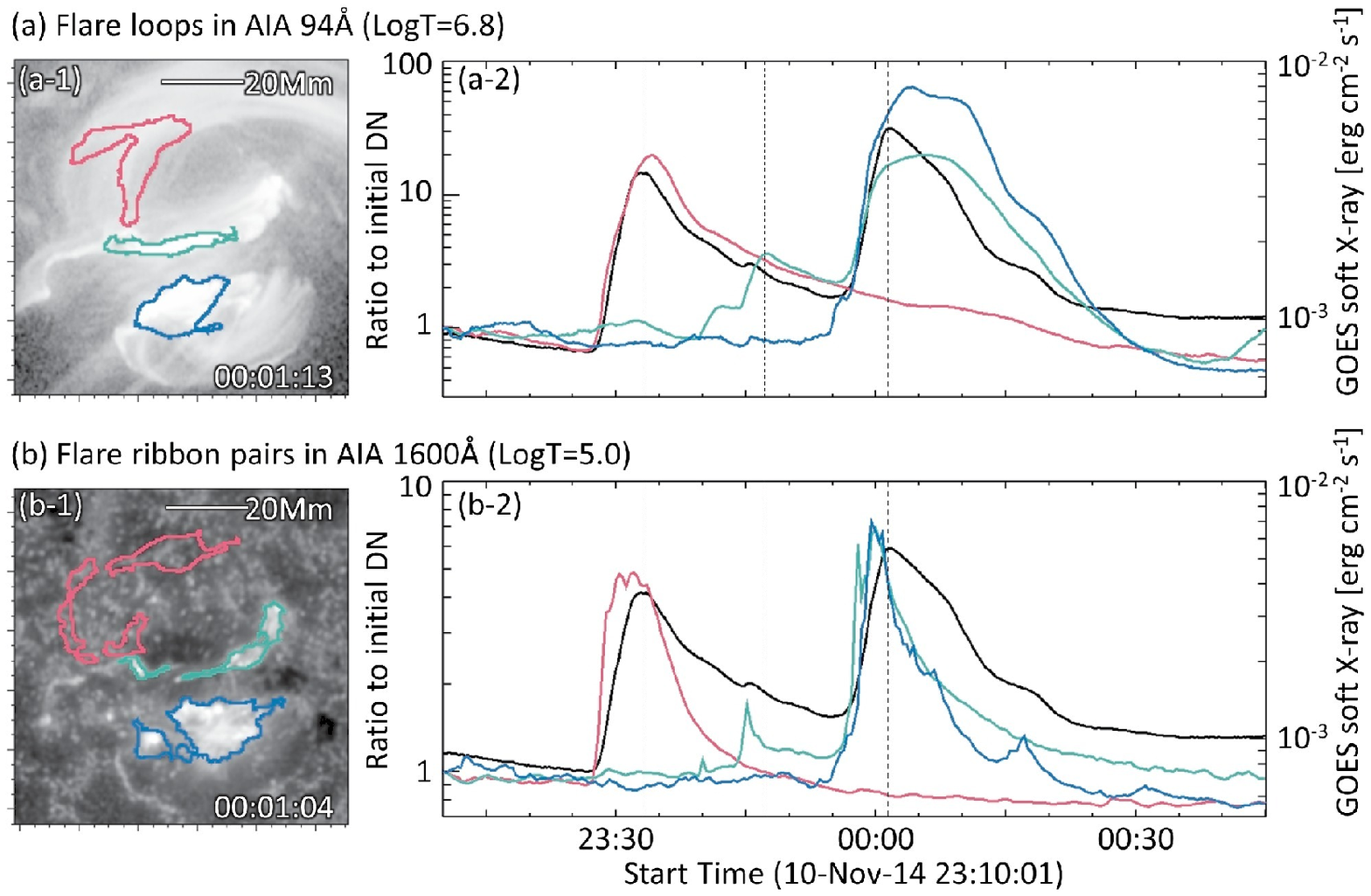}
\caption{The light curve of each flare loop or flare ribbon pair. The panels (a-1) and (b-1) are the snapshots of AIA 94\AA\ and AIA 1600\AA\ at the peak time of the latter flare, and their fields of view are the same as those of close-up views in Fig.\,\ref{fig:events_rb_fl}. The purple, green, blue contours in panel (a-1) represent the areas crossed by the northern, middle, and southern flare loops during the flares. Those in panel (b-1) are defined with the northern, middle, and southern flare ribbon pairs as well. The purple, green and blue curves in panels (a-2) and (b-2) are the temporal variations of total intensity over the region within the respective-colored contours in panels (a-1) and (b-1). These light curves are normalized with the initial values. The black solid curves are the soft X-ray flux observed by GOES. The dash lines indicate the times of panels (a), (b) and (c) of Fig.\,\ref{fig:events_aia094}--\ref{fig:events_hmimag}, and the thick ones among them also correspond to the time of panels (a-1) and (b-1).}
\label{fig:loop_ribbon}
\FigureFile(115mm,60mm){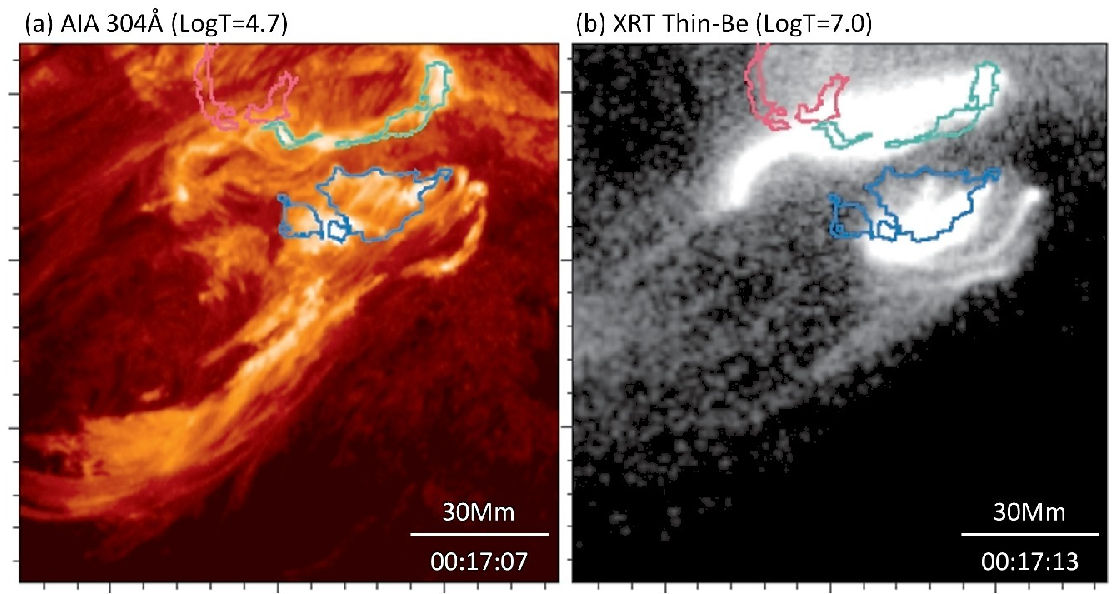}
\caption{(a) The cool jet in AIA 304\AA\ image. (b) The hot jet in XRT image. They were the simultaneous snapshots at the same time as the Fig.\,\ref{fig:jet_multi}. The colored contours correspond to those in panel (b-1) of Fig.\,\ref{fig:loop_ribbon}. The southern flare was associated with the jet ejection, which is another distinction between the southern and middle flares. The observed cool and hot jets emanated from the western periphery of the southern flare ribbon.}
\label{fig:FR_jet}
\end{center}
\end{figure*}

\begin{figure*}[tbp]
\begin{center}
\FigureFile(140mm,80mm){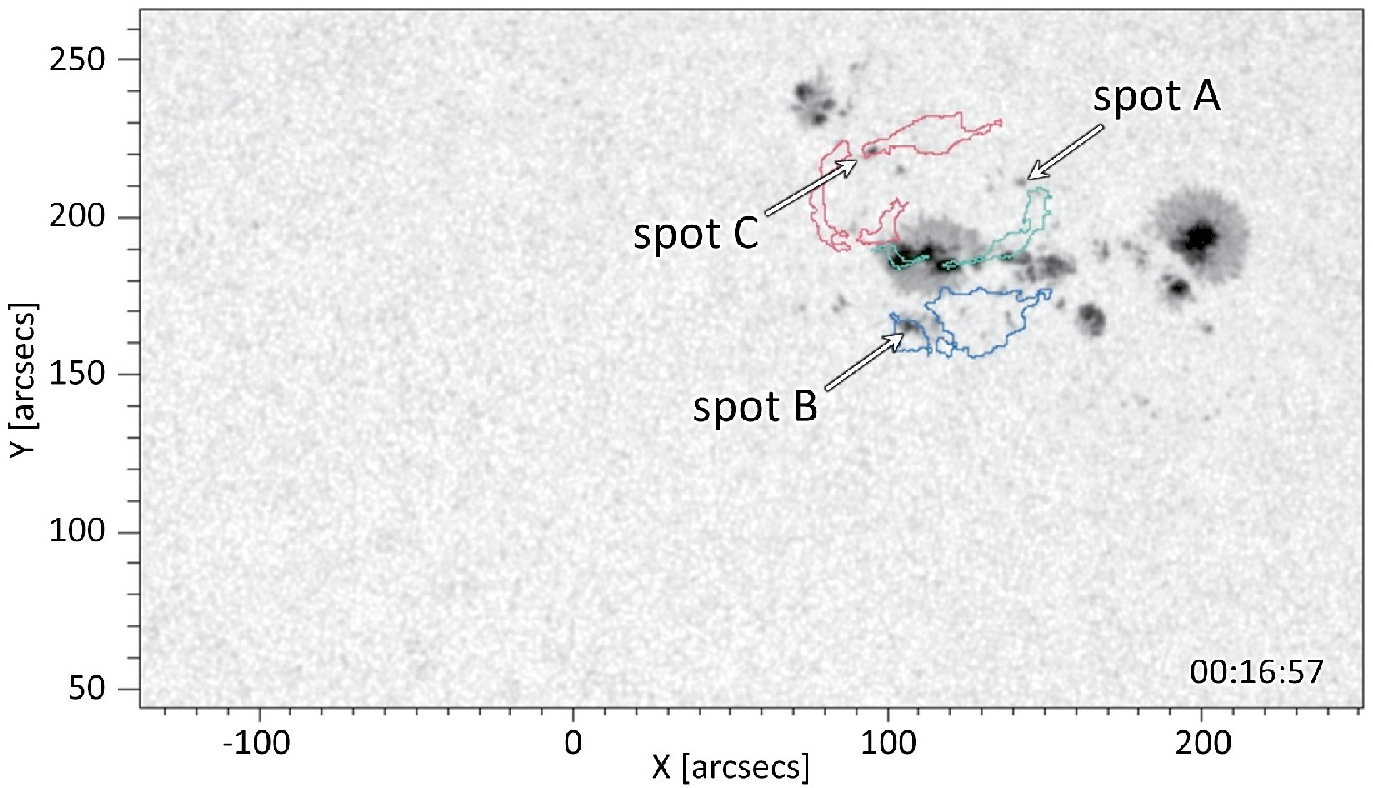}
\caption{HMI continuum image, whose field of view is the same as that of Fig\,\ref{fig:events_aia094}--\ref{fig:events_hmimag}. The colored contours correspond to those in panel (b-1) of Fig.\,\ref{fig:loop_ribbon}. The southern flare ribbon extended over a satellite spot labeled spot B, and that the other two satellite spots, labeled spot C and A, lay in the vicinity of the northern and middle flare ribbon, as well.}
\label{fig:FR_hmicon}
\end{center}
\end{figure*}

\section{Observation}
\label{sec:Observation}
Fig.\,\ref{fig:jet_multi} shows the simultaneous snapshots of the observed jet in the 304\AA\ channel of AIA, {\it Hinode}/XRT \citep{2007SoPh..243...63G} images and HMI line-of-sight magnetogram.
The multi-wavelength observation of AIA and XRT indicates that this jet consisted of the cool ($\sim10^{4.7}$K) and hot ($\sim10^{7.0}$K) components.
This multi-thermal nature of the jet will be discussed in Paper II in more detail. In the following two subsections, we report the observed features during the energy build-up and event trigger phases. \par

\subsection{Successive Flares}
\label{obs. successive flares}
In terms of the event trigger of this jet
phenomenon,
the most remarkable feature is that it emanated in the duration of the successive flares.
In fact, the time variation of soft X-ray flux observed by {\it Geostationary Operational Environmental Satellite} (GOES) shows two peaks during the observed phenomena (Fig.\,\ref{fig:goes_light_curve}), which correspond to the successive flares (C3.6 and C5.4 class) in the same active region; NOAA 12205. The observed jet was associated with the latter flare (C5.4), in particular.
The white arrow in Fig.\,\ref{fig:goes_light_curve} corresponds to the time of the snapshots in Fig.\,\ref{fig:jet_multi}.\par 

The overviews of these successive flares are presented in Fig.\,\ref{fig:events_aia094}--\ref{fig:events_hmimag}, which are the time sequence of AIA 94\AA, AIA 1600\AA\ images and HMI line-of-sight magnetograms. Fig.\,\ref{fig:events_rb_fl} is the close-up views of the black frames in Fig.\,\ref{fig:events_aia094}--\ref{fig:events_hmimag}. The times of those images are indicated with the black arrows in Fig.\,\ref{fig:goes_light_curve}. 
The red contours in those images represent the brightening features in AIA 94Å images, where the intensity was enhanced by a factor of 100 or more relative to the background level (the mean value in the field of view before the events occurrence). The yellow contours represent those in AIA 1600Å images as well, where the intensity was enhanced by a factor of 5 or more relative to the background level.
Some of those red or yellow contours respectively correspond to the flare loops (labeled FL$_{\rm N}$, FL$_{\rm S}$, FL$_{\rm M}$) or flare ribbon pairs (labeled FR$_{\rm N}$, FR$_{\rm S}$, FR$_{\rm M}$).

As seen in these images, the observed explosive phenomena involved the triple flare loop-ribbon configurations, one of which (FL$_{\rm N}$-FR$_{\rm N}$) was involved in the former flare (C3.6 class; see Fig.\,\ref{fig:goes_light_curve}), and the other two (FL$_{\rm M}$-FR$_{\rm M}$ and FL$_{\rm S}$-FR$_{\rm S}$) were in the latter flare (C5.4 class). These configurations lay in the north-south direction, and hereafter referred to as the northern, middle, and southern one. \par

It is more clearly demonstrated that the latter flare involved the two distinct magnetic configurations (middle and southern one) by paying attention to the light curve of each flare loop or flare ribbon pair, which is provided in panels (a-2) and (b-2) of Fig.\,\ref{fig:loop_ribbon}. The panels (a-1) and (b-1) in that figure are the snapshots of AIA 94\AA\ and AIA 1600\AA\ at the peak time of the latter flare, and their fields of view are the same as those of close-up views in Fig.\,\ref{fig:events_rb_fl}. The purple, green and blue curves in panels (a-2) and (b-2) are the temporal variations of total intensity over the region within the respective-colored contours in panels (a-1) and (b-1). Note that these light curves are normalized with the initial values. The black solid curves are the soft X-ray flux observed by GOES. The dash lines indicate the times of panels (a), (b) and (c) of Fig.\,\ref{fig:events_aia094}--\ref{fig:events_hmimag}, and the thick ones among them also correspond to the time of panels (a-1) and (b-1). The contours in panel (a-1) represent the areas crossed by the northern, middle, and southern flare loops during the flares, which are defined in the same way as the brightening features in Fig.\,\ref{fig:events_aia094}. The contours in panel (b-1) are defined with the northern, middle, and southern flare ribbon pairs as well, but the areas which did not brighten for longer than two minutes are removed to avoid the bright ejecta in AIA 1600\AA\ images. \par

As seen in Fig.\,\ref{fig:loop_ribbon}, the light curves of middle flare loop and ribbon pairs are different from those of southern ones in the existence of transient small brightening around 23:45UT. The southern flare did not show this pre-flare emission, and suddenly brightened along with the main phase of the middle flare.
Moreover, the southern flare was associated with the jet ejection, which is another distinction between the southern and middle flares. Fig.\,\ref{fig:FR_jet} shows that the observed cool and hot jets emanated from the western periphery of the southern flare ribbon. The colored contours in that figure are the same as those in panel (b-1) of Fig.\,\ref{fig:loop_ribbon}.\par
The above triple successive flares possibly mean that the magnetic energy for the observed explosive phenomena had been separately stored around each flare sites. To investigate such an individual energy build-up process, we pay particular attention to the three satellite spots seen in Fig.\,\ref{fig:FR_hmicon}. In that figure, it is shown that the northern, middle, and southern flare ribbon pairs extended over the satellite spots labeled spot C, A, and B, respectively. That means the magnetic flux of these satellite spots were involved in the reconnection. In the next subsection, therefore, we discuss the temporal evolution of these three satellite spots in terms of the energy build-up process of the triple successive flares.


\subsection{Emergence of Satellite Spots}
\begin{figure*}[tb]
\begin{center}
\FigureFile(170mm,85mm){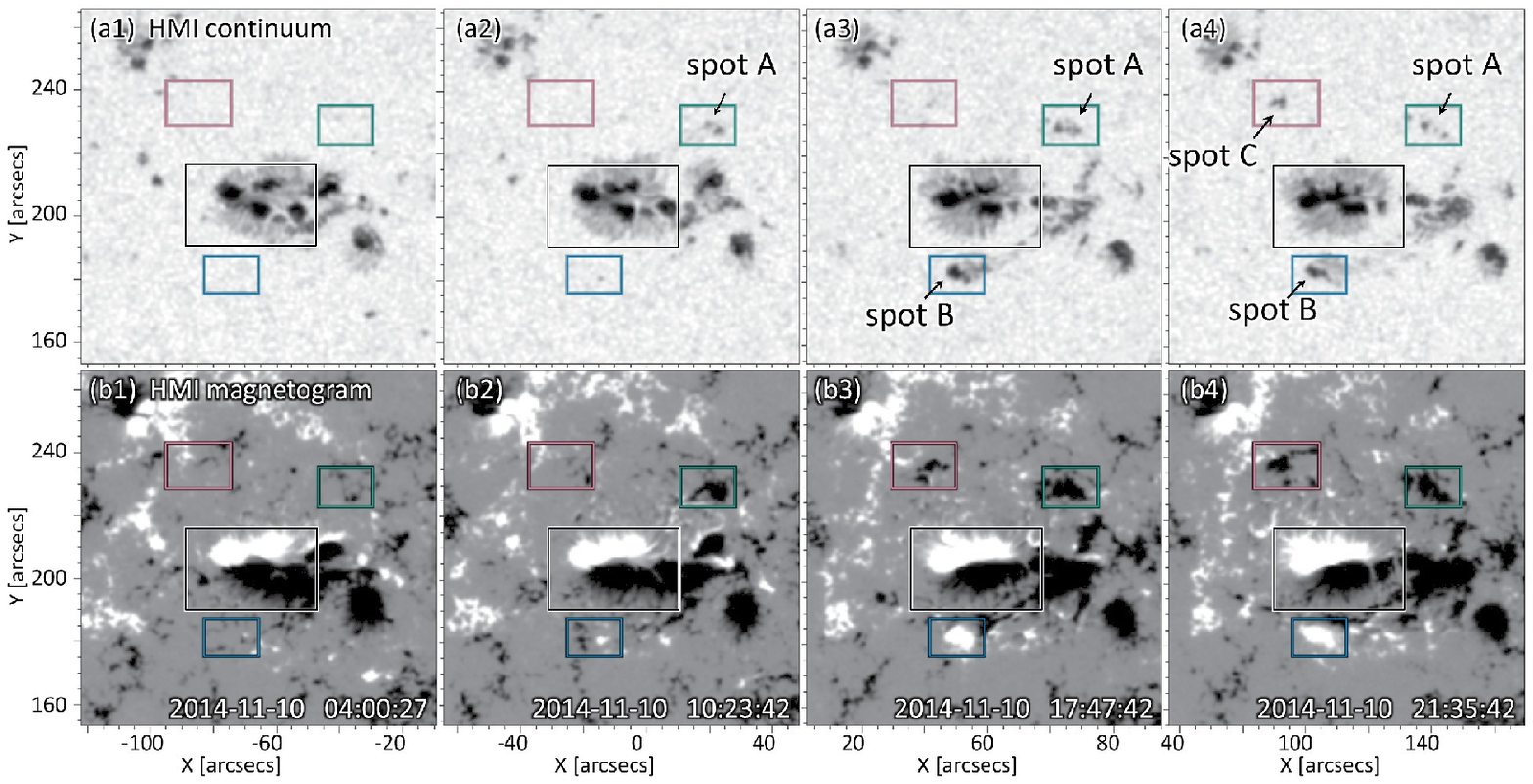}
\caption{The morphological evolutions of the satellite spots. The green, blue, and purple frames indicate the emerging regions of the spot A, B, and C. }
\label{fig:sat_birth}
\FigureFile(110mm,80mm){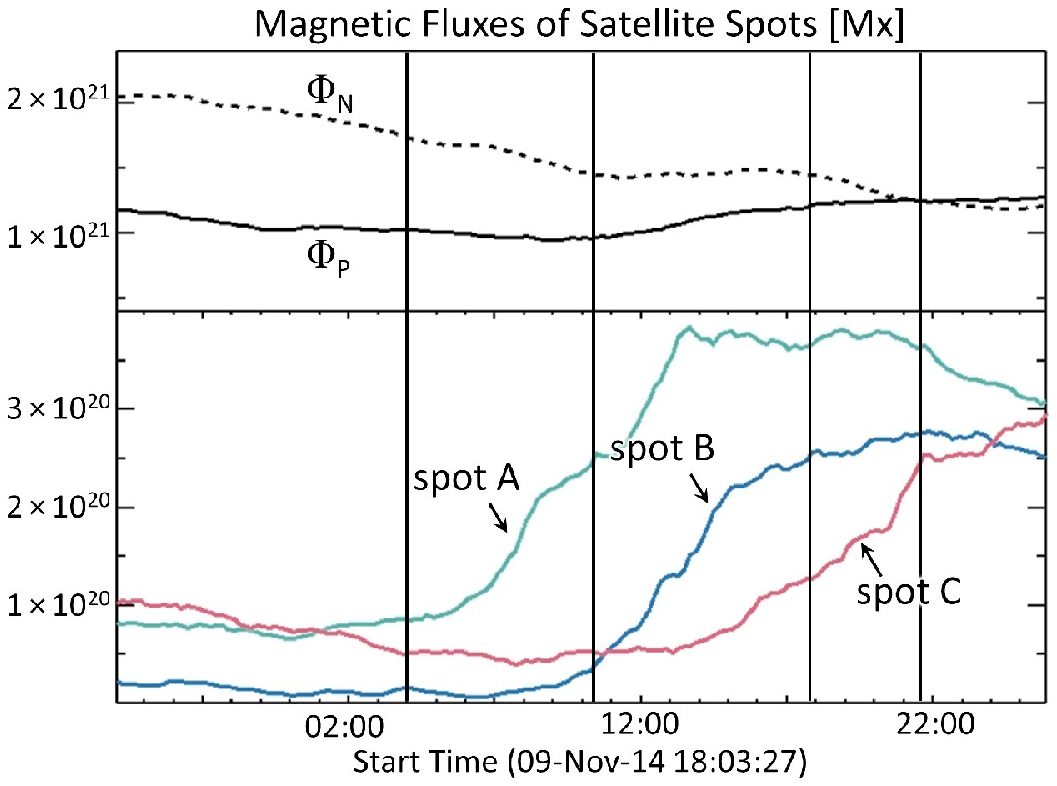}
\caption{The temporal variation of the total positive magnetic flux within the blue frames, and the negative magnetic flux within the purple and green frames, along with that of negative (dash curve) and positive (solid curve) magnetic fluxes around the pivotal $\delta$-type sunspot within the black frames in Fig.\,\ref{fig:sat_birth}.. The vertical lines in that figure correspond to the times of images in Fig.\,\ref{fig:sat_birth}.}
\label{fig:sat_flux}
\end{center}
\end{figure*}

The morphological evolutions of the satellite spots are presented in Fig.\,\ref{fig:sat_birth}, where the green, blue, and purple frames indicate the emerging regions of the spot A, B, and C. 
Fig.\,\ref{fig:sat_flux} shows the absolute value of the total fluxes of these satellite spots, along with the negative (dash curve) and positive (solid curve) magnetic fluxes around the pivotal $\delta$-type sunspot within the black frames in Fig.\,\ref{fig:sat_birth}.
Note that the vertical lines in that figure correspond to the times of images in Fig.\,\ref{fig:sat_birth}. As seen in these figures, the three satellite spots had emerged in order of spot A, B, C during 20 hours prior to the event occurrence. \par

\begin{figure*}[tbp]
\begin{center}
\FigureFile(170mm,118mm){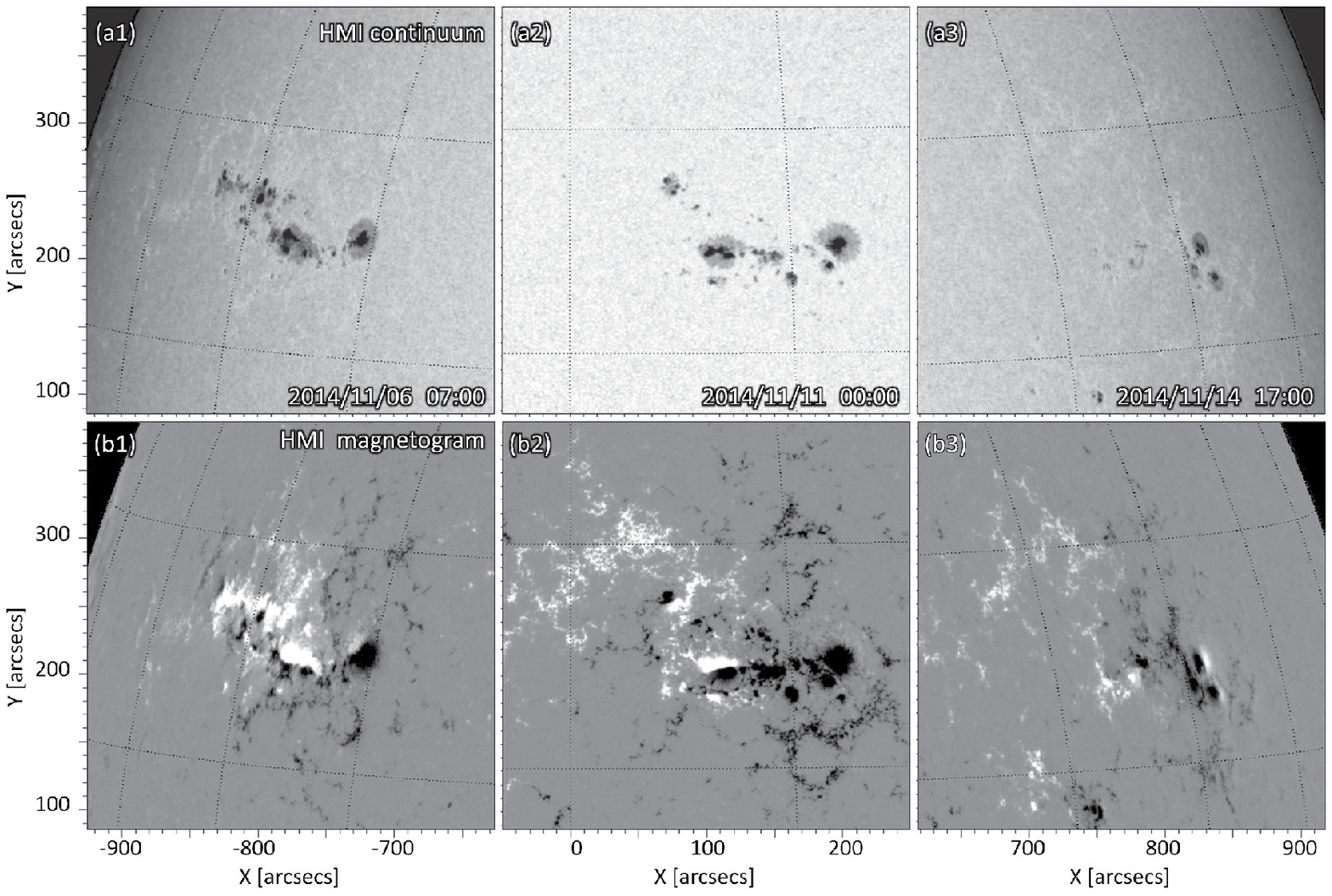}
\caption{The evolution of the entire active region NOAA 12205. This active region had already retained the complex magnetic configuration at the time of its appearance from the east solar limb, and was categorized into the $\beta\gamma\delta$-type at that time.}
\label{fig:ev_NOAA12205}
\FigureFile(80mm,80mm){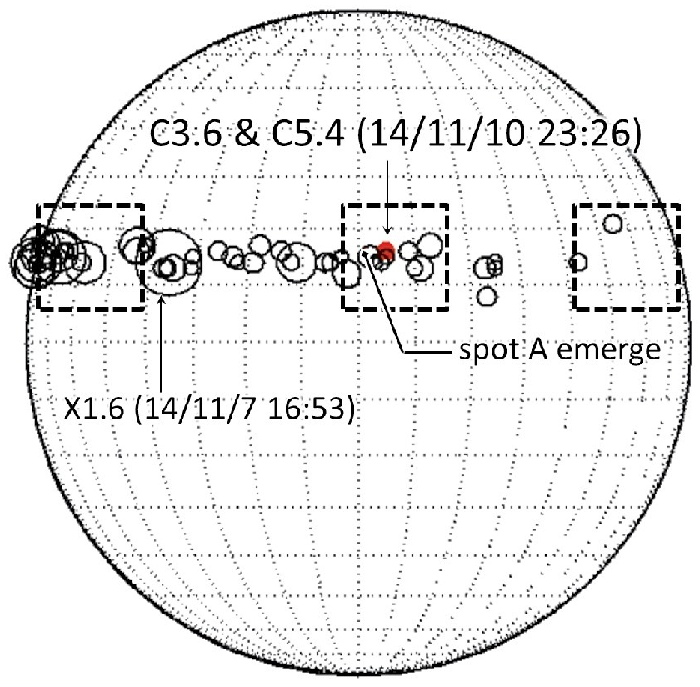}
\caption{The major flares in the active region NOAA12205 which are listed in the {\it SolarSoft} Latest Events Archive. The circles represent the sites of flares, and their sizes correspond to the maximum magnitude of the soft X-ray flux observed by GOES. The observed flares are indicated with a filled red circle. The panels of Fig.\,\ref{fig:ev_NOAA12205} are the white light images and the line-of-sight magnetograms of the dash-line frames in this figure.}
\label{fig:flare_NOAA12205}
\end{center}
\end{figure*}

These satellite spots' emergence appeared to be associated with the disintegration of the pivotal $\delta$-type sunspot.
The number of the umbra patches within one penumbra in Fig.\,\ref{fig:sat_birth} decreased from 5 to 3, and Fig.\,\ref{fig:sat_flux} shows that about a half of the negative magnetic flux in the black frames of Fig.\,\ref{fig:sat_birth} had disappeared in 30 hours.
In addition, the entire active region NOAA 12205 was in the decaying phase at that time. This active region had already retained the complex magnetic configuration at the time of its appearance from the east solar limb, and was categorized into the $\beta\gamma\delta$-type at that time, according to the Mount Wilson magnetic classification \citep{1919ApJ....49..153H} (see panels (a1), (b1) of Fig.\,\ref{fig:ev_NOAA12205}). That means this active region consisted of the bipolar without the unique boundary between the opposite polarities ($\beta\gamma$-type sunspot group), and the sunspot with strongly sheared magnetic field in itself ($\delta$-type sunspot). Such a complex magnetic configuration had brought about the many major flares in this region up to X1.6 class flare \citep{2016A&A...596A...1S}, as shown in Fig.\,\ref{fig:flare_NOAA12205}. In that figure, the circles represent the sites of flares listed in the {\it SolarSoft} Latest Events Archive (http://www.lmsal.com/solarsoft/latest\_events\_archive.html), and their sizes correspond to the maximum magnitude of the soft X-ray flux observed by GOES. The observed flares are indicated with a filled red circle. The panels of Fig.\,\ref{fig:ev_NOAA12205} are the white light images and the line-of-sight magnetograms of the dash-line frames in Fig.\,\ref{fig:flare_NOAA12205}. These figures demonstrate that both the flare activity and magnetic configuration of this region were decaying during the period of interest in this study, from the satellite spots' emergence (indicated in Fig.\,\ref{fig:flare_NOAA12205}) to the explosive phenomena. \par
Finally in this section, we summarize the observed features during the energy build-up phase and event trigger phase, as follows:\\
(i) The observed explosive phenomena involved the triple flare loop-ribbon configurations which lay in the north-south direction. The successive flares started with the northern flare, and followed the middle and southern flares. It is remarkable that the pre-flare emission from the middle configuration was observed 15 minutes before the latter flares,
 and that the jet ejection was associated with the southern flare.\\
(ii) The above three pairs of flare ribbon extended to the vicinity of the three satellite spots, respectively. They had emerged around the $\delta$-type sunspot in the decaying active region during 20 hours prior to the event occurrence.\par

\section{Analysis \& Results}
\label{sec:Analysis}

We analyzed the temporal evolutions of the photospheric horizontal flow field around the satellite spots by using the optical flow method named Nonlinear Affine Velocity Estimator (NAVE; \cite{2008ApJ...689..593C}). In this method, the velocity vector of each pixel is determined by estimating the flow field in a given local area around the pixel in the form of the affine transformation. We applied the method to the time series of HMI line-of-sight magnetograms with a cadence of three minutes, and defined the spatial scale of the local area as 9 arcseconds. Note that the flow field resulting from the solar rotation  is reduced in advance of this analysis, by using \texttt{drot\_map.pro} incorporated in {\it SolarSoft}.\par
\begin{figure*}[tbp]
\begin{center}
\FigureFile(136mm,228mm){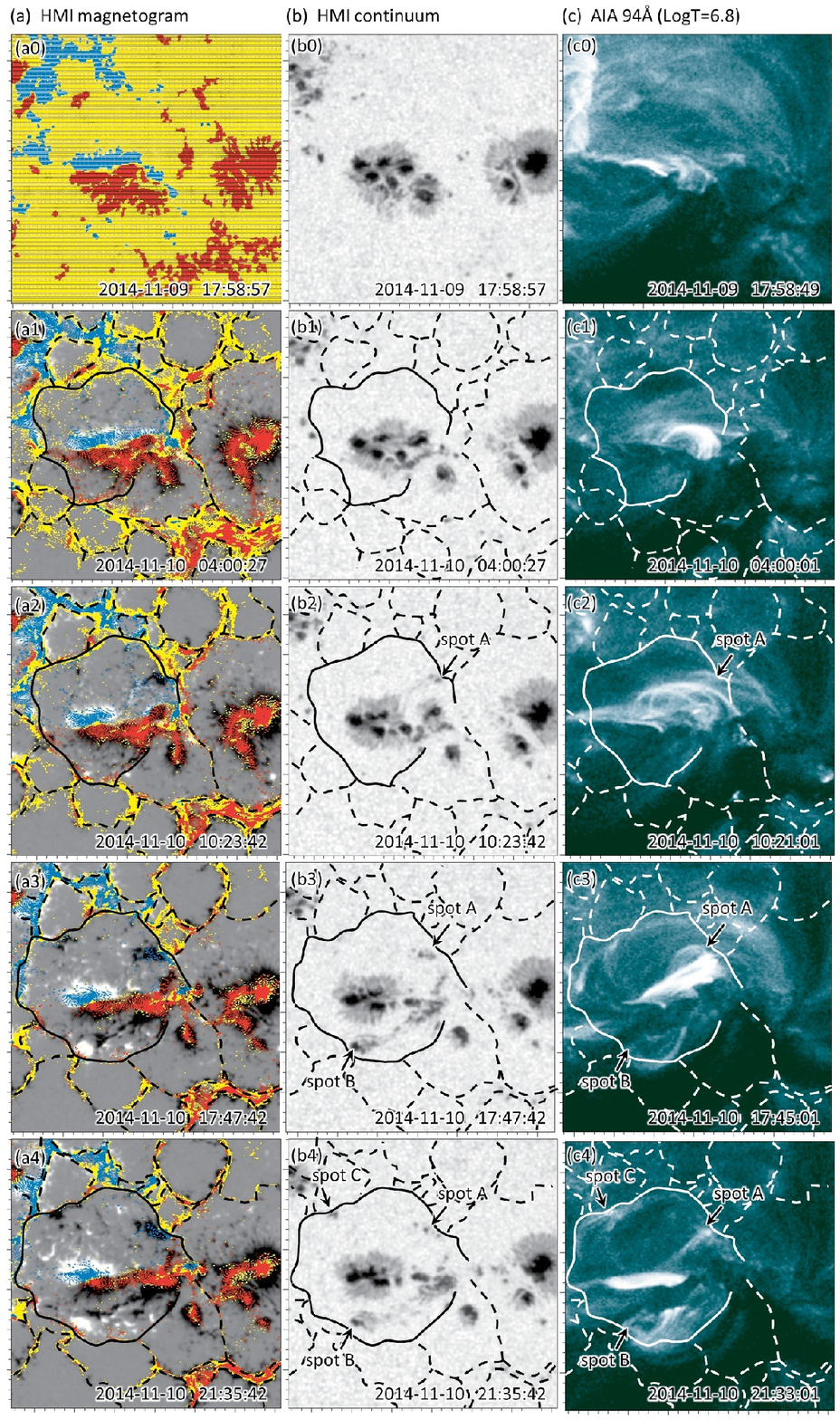}
\caption{The horizontal flow field on the photosphere. The panels (a1)--(a4) are the HMI line-of-sight magnetograms at the same times as those of Fig.\,\ref{fig:sat_birth}, and overlaid with the Lagrange particles which are evenly scattered 30 hours before the event occurrence (see panel (a0)). The blue and red particles represent the magnetic elements with the magnetic strength over 200 G when they are scattered. The convergent boundaries are indicated with the solid or dash lines in panels (a1)--(a4), and drawn over the time-sequence of HMI continuum (panels (b1)--(b4)). The solid lines enclose the moat region of the $\delta$-type sunspot.}
\label{fig:ar_ev}
\end{center}
\end{figure*}

Fig.\,\ref{fig:ar_ev} is the analysis result. The panels (a1)--(a4) of that figure are the HMI line-of-sight magnetograms at the same times as those of Fig.\,\ref{fig:sat_birth}, and overlaid with the Lagrange particles which are evenly scattered 30 hours before the event occurrence (see panel (a0)). The blue and red particles represent the magnetic elements with the magnetic strength over 200 G when they are scattered.
Note that the newly emerging field does not appear as the particles in that figure but affect the pre-existed particles by its divergent flow.
\par
As a result of those Lagrange particles carried by the photospheric flow, the network structure appears to be formed promptly (see panel (a1) of Fig.\,\ref{fig:ar_ev}). This result clearly demonstrates that the photospheric horizontal flow field is consisted of many tectonic-plates-like territories, in each of which an inherent divergent flow is dominant, with flux emergence, moat flow, or supergranulation. Such convergent boundaries are indicated with the solid or dash lines in panels (a1)--(a4), and drawn over the time-sequence of HMI continuum (panels (b1)--(b4)) and AIA 94\AA\ images (panels (c1)--(c4)). The solid lines enclose the moat region of the $\delta$-type sunspot.

The most remarkable feature in Fig.\,\ref{fig:ar_ev} is that the all of the three satellite spots emerged in the moat region of the $\delta$-type sunspot, especially near its convergent boundary with the neighboring supergranules or moat regions of adjacent sunspots.
AIA 94\AA\ images show that the bright coronal loops of the satellite spots successively emerged around the pivotal $\delta$-type sunspot, which suggests the continuous energy build-up in the corona, with the magnetic configuration confined in one large convection cell.
\par

\begin{figure*}[tbp]
\begin{center}
\FigureFile(170mm,124mm){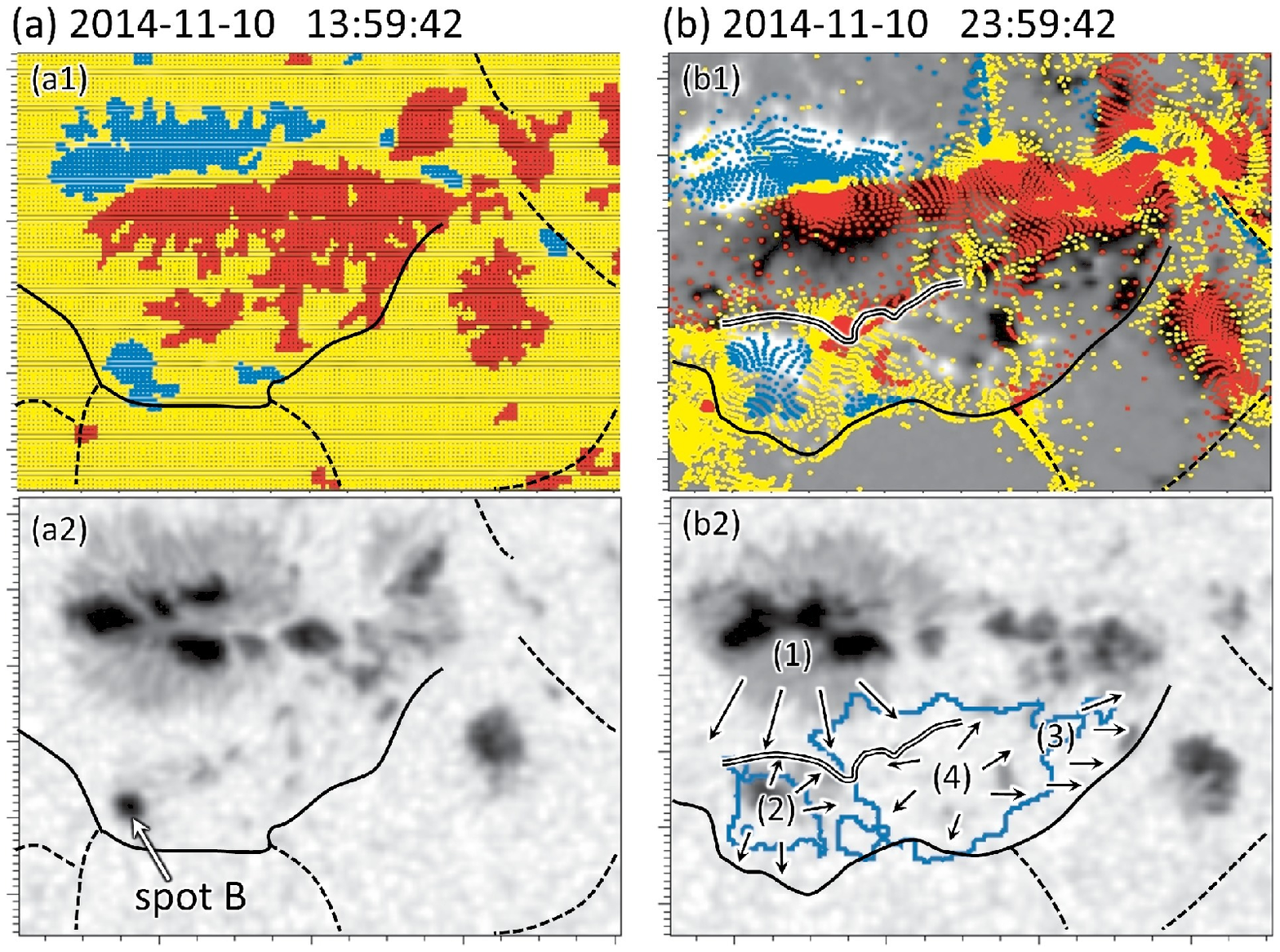}
\caption{The close-up views of Fig.\,\ref{fig:ar_ev} around the spot B, but the Lagrange particles are scattered after the spot B emergence (10 hours before the event occurrence; see panels (a1), (a2)). The solid and dash lines in panels (a1), (a2) correspond to the convergent boundaries of the Lagrange particles scattered at 17:58UT on 2014/11/09 (the same time as panels (a0), (b0) in Fig.\,\ref{fig:ar_ev}). The blue contours in panel (b2) represent the southern flare ribbon (see Fig.\,\ref{fig:FR_hmicon}), from the western periphery of which the observed jet emanated. The double lines in panels (b1), (b2) represent the polarity inversion line along the convergent boundaries of the photospheric flow field.}
\label{fig:flow_spot}
\end{center}
\end{figure*}

Here, we pay particular attention to the horizontal flow field around the spot B after its emergence, for the discussion in Paper II on the relationship between the properties of the jet and the magnetic configuration around its ejection site. Fig.\,\ref{fig:flow_spot} is the close-up views of Fig.\,\ref{fig:ar_ev} around the spot B, but the Lagrange particles in that figure are scattered after the spot B emergence (10 hours before the event occurrence; see panels (a1), (a2)). The solid and dash lines in panels (a1), (a2) correspond to the convergent boundaries of the Lagrange particles scattered at 17:58UT on 2014/11/09 (the same time as panels (a0), (b0) in Fig.\,\ref{fig:ar_ev}). The blue contours in panel (b2) represent the southern flare ribbon (see Fig.\,\ref{fig:FR_hmicon}), from the western periphery of which the observed jet emanated. \par
The notable feature in this figure is the fragmentation of the area which had been originally the moat region of the $\delta$-type sunspot until the spot B emergence. The horizontal flow field in that area became composed of the several flow channels; (1) the moat flow from the pivotal $\delta$-type sunspot, (2) that from the satellite spot B, (3) the proper motion of the leading polarities of the spot B, and (4) the divergent flow associated with the flux emergence into the mid of the magnetic configuration of the spot B. As shown by the contours of the southern flare ribbon, both of the satellite spot B and the emerging flux were involved in the southern flare and jet ejection.  In fact, since their positive polarities encountered the global magnetic field with negative polarity in the moat region of the $\delta$-type sunspot, the polarity inversion line was naturally formed along the convergent boundary between the flow channels (1) and (2), (4), which is indicated with the double lines in panels (b1), (b2).


\section{Discussion}
\label{sec:Discussion}
In this section, we discuss the relationship between the satellite spots' emergence and the observed explosive phenomena, including the successive flares and jet ejection. The successive flares involved the triple flare loop-ribbon configurations, each of which connected to the satellite spot, respectively. All of the three satellite spots emerged in the moat region of the pivotal $\delta$-type sunspot, especially near its convergent boundary with the neighboring supergranules or moat regions of adjacent sunspots.
That suggests the continuous energy build-up in the corona, with the magnetic configuration confined in one large convection cell.
\par
\begin{figure*}[p]
\begin{center}
\FigureFile(142mm,200mm){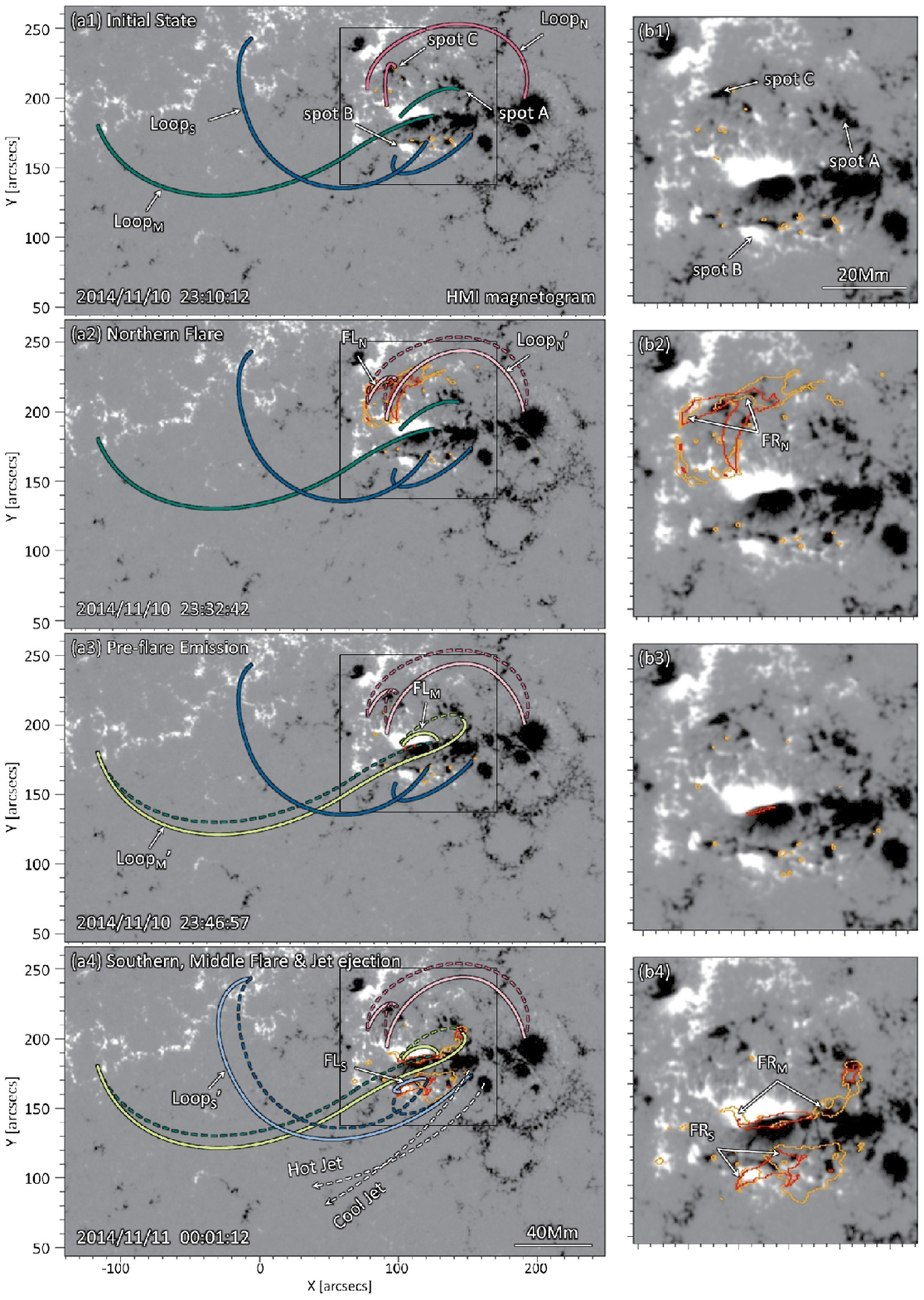}
\caption{The temporal evolution of the magnetic configuration during the observed explosive phenomena. The background images are the time-sequence of HMI line-of-sight magnetograms, whose fields of view are the same as those of Fig.\,\ref{fig:events_aia094}--\ref{fig:events_hmimag}. The red or yellow contours represent the flare loops and ribbons, as well as Fig.\,\ref{fig:events_aia094}--\ref{fig:events_hmimag}. The close-up views of black frames in panels (a1)--(a4) are presented in panels (b1)--(b4). At the initial state (panel (a1)), the coronal field around the $\delta$-type sunspot included the three magnetic field systems. Each configuration of them is composed of the magnetic flux of one satellite spot A, B or C and its pre-exiting ambient field Loop$_{\rm M}$, Loop$_{\rm S}$, or Loop$_{\rm N}$., as indicated with the pairs of dark green, blue or purple lines. Through the northern flare, the magnetic flux of spot C and Loop$_{\rm N}$ (dark purple lines in panel (a)) reconnected to the flare loop FL$_{\rm N}$ and Loop$_{\rm N}$' (light purple lines in panel (b)), while the flare ribbon FL$_{\rm N}$ brightened at the footpoints of FL$_{\rm N}$. After that, the transient small brightening occurred in the middle system (Fig.\,\ref{fig:loop_ribbon}), which was composed of the magnetic flux of the spot A and Loop$_{\rm N}$ (dark green lines in panel (a), (b)). The energy release proceeded with the reconnection between them to the pair of FL$_{\rm M}$ and Loop$_{\rm M}$' (light green lines in panel (a3). This small energy release of the middle flare had been hindered by the overlying Loop$_{\rm S}$ of the southern configuration during the pre-flare emission phase. This restriction was finally removed due to the reconnection between Loop$_{\rm S}$\ and the magnetic flux of satellite spot B.}
\label{fig:coronal_field}
\end{center}
\end{figure*}

From these results, we conjecture the scenario, in which a jet ejects during the successive flares which involves the triple flare loop-ribbon configurations. Fig.\,\ref{fig:coronal_field} is the schematic drawing which shows the temporal evolution of the coronal field during the successive flares. The background images are the time-sequence of HMI line-of-sight magnetograms. Their fields of view are the same as those of Fig.\,\ref{fig:events_aia094}--\ref{fig:events_hmimag}, and the red or yellow contours represent the flare loops and ribbons, as well as Fig.\,\ref{fig:events_aia094}--\ref{fig:events_hmimag}. The close-up views of black frames in panels (a1)--(a4) are presented in panels (b1)--(b4).\par

At the initial state (panel (a1)), the coronal field around the $\delta$-type sunspot included the three magnetic field systems. Each configuration of them is composed of the magnetic flux of one satellite spot A, B or C and its pre-exiting ambient field Loop$_{\rm M}$, Loop$_{\rm S}$, or Loop$_{\rm N}$, as indicated with the pairs of dark green, blue or purple lines in Fig.\,\ref{fig:coronal_field}.

\begin{figure*}[tp]
\begin{center}
\FigureFile(120mm,60mm){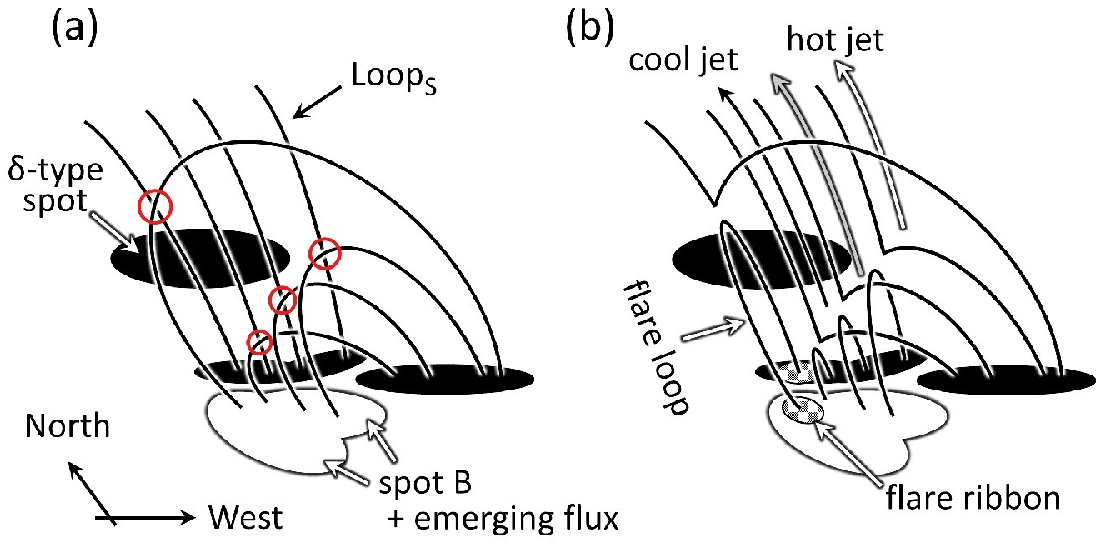}
\caption{The schematic drawing for the transformation of the magnetic configuration around the satellite spot B through the southern flare. The red circles in panel (a) point to the sites of current sheets, the altitudes of which distribute across so wide range that the resultant explosive phenomena are characterized with the various temperatures. The magnetic configuration in panel (a) had been formed, as the spot B and the adjacent emerging flux encountered the global coronal field in the moat region of the $\delta$-type sunspot.}
\label{fig:satellite_mag}
\end{center}
\end{figure*}

A series of phenomena began with the northern flare, which was triggered by the evolution of the satellite spot C (see Fig.\,\ref{fig:sat_flux}). During this flare, the magnetic flux of spot C and Loop$_{\rm N}$ (dark purple lines in panel (a)) reconnected to the flare loop FL$_{\rm N}$ and Loop$_{\rm N}$' (light purple lines in panel (b)), while the flare ribbon FL$_{\rm N}$ brightened at the footpoints of FL$_{\rm N}$.\par

The first sign of the explosive phenomena subsequent to the northern flare appeared as the transient small brightening from the middle system (Fig.\,\ref{fig:loop_ribbon}), which was composed of the magnetic flux of the spot A and Loop$_{\rm M}$ (dark green lines in panel (a), (b)). The energy release proceeded with the reconnection between them to the pair of FL$_{\rm M}$ and Loop$_{\rm M}$' (light green lines in panel (a3). This pre-flare emission possibly resulted from the destabilization of the middle system due to the change of the northern magnetic configuration, although the detailed process is unknown in our study. On the other hand, as stressed in section \ref{sec:Analysis}, this small energy release process immediately ceased without monotonically leading to the main phase of the middle flare, and it was not seen from the southern magnetic configuration. One possibility accounting for this fact is that the energy release of the middle flare had been hindered by the overlying Loop$_{\rm S}$ of the southern configuration during the pre-flare emission phase. This restriction was finally removed due to the reconnection between Loop$_{\rm S}$ and the magnetic flux of satellite spot B, and thus the main phase of the middle flare was triggered at the same time as the onset of the southern flare. Therefore, the energy build-up process around the spot B played a remarkable role in the trigger of the middle, southern flares. \par

In addition to these flares, the magnetic configuration around the spot B was responsible for the jet ejection with the various characteristic temperature from $10^4$K to $10^7$K (see Fig.\,\ref{fig:jet_multi}). Fig.\,\ref{fig:satellite_mag} is the schematic image for the magnetic configuration around the spot B. The red circles in panel (a) of that figure correspond to the reconnection sites, leading to the southern flare and jet ejection. According to the unified model (section \ref{sec:Introduction}), the relative height of the energy release site (reconnection site) to the transition layer determines the properties of the associated explosive phenomena. In the case of the observed phenomena, the footpoint of the cool jet was to the west of the southern flare loop and to the east of that of hot jet (see Fig.\,\ref{fig:jet_multi}). This suggests the energy release site of the jet migrated from east to west and relatively lower atmosphere to higher atmosphere, while that of the southern flare lay in the easternmost and highest of all. This is also suggested by the horizontal flow field around the spot B after its emergence. As shown in Fig.\,\ref{fig:flow_spot}, the energy storage around the spot B is represented by the formation of the polarity inversion line, along which the positive polarities of spot B and adjacent emerging flux encountered the global magnetic field with negative polarity (Loop$_{\rm S}$) in the moat region of the pivotal $\delta$-type sunspot. Thus, it is suggested that the current sheets were formed at the various altitudes from the lower atmosphere, where the newly emerging flux contacted with the Loop$_{\rm S}$, to the upper atmosphere, where the fully evolved satellite spot B did. \par

Finally, we refer to the characteristics of the satellite spots' emergence which led to the observed explosive phenomena. It is most remarkable that all of these spots emerged in the moat region of the pivotal $\delta$-type sunspot, especially near its convergent boundaries with the neighboring supergranules or moat regions of the adjacent sunspots, during the decay phase of the entire active region. Such characteristics possibly suggest that their magnetic fluxes had been originally incorporated in the flux tube of $\delta$-type sunspot and were detached along with its disintegration, so that the moat flow advected them to the convergent boundaries among the larger convection cells before they emerged above the photosphere. This study is not able to answer why the detached magnetic fluxes were led to emerge after they reach the convergent boundaries. Recent numerical experiments (e.g., \cite{2015ApJ...814..125R}) address the various nature of the moat flow; its origin, connection to the penumbra and Evershed flow, or role in decaying a sunspot. There are also the observational studies on this issue in terms of both the statistical approach \citep{2013A&A...551A.105L} and case study \citep{2007ApJ...671.1013D}. Because these previous researches focus on the isolated sunspots, further studies are needed to investigate what makes the detached magnetic fluxes emerge around the pre-exiting the sunspot as the satellite spots.

\section{Summary}
\label{sec:Summary}
We observed a solar jet phenomenon associated with successive flares on November 10th 2014. These explosive phenomena were involved with the satellite spots' emergence around the $\delta$-type sunspot in the decaying active region NOAA 12205. In order to investigate the energy build-up process of the observed phenomena, we analyzed the flow field on the photosphere with the optical flow method. The observational and analysis results are summarized as follows;\\
(i) The observed explosive phenomena involved three satellite spots, which successively reconnected with their pre-existing ambient fields.\\
(ii) All of these satellite spots emerged in the moat region of a pivotal $\delta$-type sunspot, especially near its convergent boundary with the neighboring supergranules or moat regions of adjacent sunspots.\\
(iii) Around the jet ejection site, the positive polarities of satellite spot and adjacent emerging flux encountered the global magnetic field with negative polarity in the moat region of the pivotal $\delta$-type sunspot, and thus the polarity inversion line was formed along the convergent boundary among the photospheric horizontal flow channels.
\\
\par
\begin{ack}
{\it Hinode} is a Japanese mission developed and launched by ISAS/JAXA, with NAOJ as domestic partner and NASA and UKSA as international partners. It is operated by these agencies in co-operation with ESA and NSC (Norway). HMI and AIA are instruments on board SDO, a mission for NASA's Living With a Star program. IRIS is a NASA small
explorer mission developed and operated by LMSAL with mission operations executed at NASA Ames Research center and major contributions to downlink communications funded by ESA and the Norwegian Space Centre. This work was supported by JSPS KAKENHI Grant Numbers JP16H03955, JP15H05814, and JP15K17772. AA is supported by a Shiseido Female Researcher Science Grant. 
\end{ack}

\end{document}